\documentclass[prd,showpacs,preprintnumbers,floatfix, two column]{revtex4}
\usepackage{amsmath}
\usepackage{graphicx}
\begin{document}
\title{Entanglement Entropy: Helicity versus Spin}
\date{\today }
\author{Song He}

 \affiliation{Institute of Theoretical
Physics, School of Physics, Peking University, Beijing, 100871,
China}
 \author{Shuxin Shao}

 \affiliation{Department of Physics,
Beijing Normal University, Beijing, 100875, China}
\author{Hongbao Zhang\footnote{Current address is Perimeter Institute for Theoretical Physics, Waterloo, Ontario, N2L 2Y5, Canada.}}

    \affiliation{Department of Astronomy, Beijing Normal University, Beijing, 100875,
    China\\
    Department of Physics, Beijing Normal University,
    Beijing, 100875, China\\
 CCAST (World
Laboratory), P.O. Box 8730, Beijing,
   100080, China}

\begin{abstract}
For a massive spin $\frac{1}{2}$ field, we present the reduced spin
and helicity density matrix, respectively, for the same pure one
particle state. Their relation has also been developed. Furthermore,
we calculate and compare the corresponding entanglement entropy for
spin and helicity within the same inertial reference frame. Due to
the distinct dependence on momentum degree of freedom between spin
and helicity states, the resultant helicity entropy is different
from that of spin in general. In particular, we find that both
helicity entanglement for a spin eigenstate and spin entanglement
for a right handed or left handed helicity state do not vanish and
their Von Neumann entropy has no dependence on the specific form of
momentum distribution as long as it is isotropic.
\end{abstract}

\pacs{03.67.Mn 03.65.Ge 03.65.Ud} \maketitle

Quantum information theory is usually formulated in the framework of
non-relativistic quantum mechanics, since particles moving at
relativistic speeds may not be needed to realize the promise of
quantum information process such as quantum computation. However,
relativity, especially special relativity plays a significant role
in quantum entanglement and related quantum technology, such as
teleportation. This point is obviously justified by quantum optics,
which is well established on the basis of not only quantum theory
but also special relativity in nature\cite{Walls}. For example, most
of EPR-type experiments have been performed by photon
pairs\cite{Aspect,Weihs}. In addition, experiments of quantum
teleportation have also been extensively carried out by
photons\cite{Bouwmeester,Furusawa}.

On the other hand, considerable efforts have also been expanded on
the theoretical investigation of quantum information theory in
relativistic framework, which has gone beyond from photons to
electrons, and from explicit examples calculated in some specific
cases to general framework formulated in relativistic quantum
mechanics and even relativistic quantum field
theory\cite{Czachor1,Peres1,Peres2,Alsing1,Gingrich,Peres3,Peres4,Pachos,
Enk,Czachor2,Bergou,Alsing2,Peres5,Soo,Alsing3,Shi1,Kim,Czachor3,FM,Caban1,Kok,Ball,Lamata1,Lamata2,
Jordan1,Alsing4,Shi2,Caban2,Jordan2,Ling,Adesso}. For review, please
refer to \cite{Peres5}. However, for Dirac fields, previous
discussions of relativistic quantum information theory are
restricted primarily on quantum entanglement between spin and
momentum degrees of freedom. Only recently has quantum entanglement
between helicity and momentum been systematically formulated and
numerically analyzed in relativistic framework for the first time in
\cite{He}. This pioneering work not only sets a starting point for
further worthwhile investigations along this line, but also acquires
much importance from the perspective of high energy physics. It is
the helicity rather than spin that is more often under both
theoretical consideration and experimental detection in high energy
physics, since the helicity has an advantage in providing a smooth
transition to the massless case. Although both the helicity states
and spin states can constitute the basis of Hilbert space of one
particle, they differ in the way of unitary transformation under the
action of Lorentz group. As a result, the entanglement properties
for helicity differs remarkably from those for spin after we trace
out the momentum degree of freedom. Especially, it is found that in
the sharp momentum limit, unlike the vanishing spin entropy, at
small velocities of the inertial observer the helicity entropy
demonstrates a sudden jump onto a constant value, half of the
entropy for the maximal entangled Bell states, which may be easily
observed in high energy physics experiments.

As a further step along this direction, the purpose of this paper is
to investigate the explicit difference between the reduced spin and
helicity density matrix. Especially, for a pure one particle state,
we find the resultant helicity entropy is different from that for
spin with respect to the same inertial observer.

Start with a field with positive mass $m$ and spin $\frac{1}{2}$, we
can construct the spin states $|p,\sigma\rangle$ as a complete
orthonormal basis for Hilbert space of one particle, where $p$ is
the four momentum of particle, and $\sigma$ represents the spin
along the $z$ axis. Similarly, we can also choose the helicity
states $|p;\lambda\rangle$ to form a complete orthonormal basis for
the same Hilbert space with the helicity denoted by $\lambda$. The
former is usually called spin representation, and the latter is
called helicity representation. Moreover, these spin states and
helicity states are related by the representation transformation
as\cite{Weinberg}
\begin{equation}
|p;\lambda\rangle=D_{\sigma\lambda}[R(p)]|p,\sigma\rangle.\label{TR}
\end{equation}
Here $R(p)$ is the rotation that carries the $z$ axis into the
direction $\mathbf{p}$, and $D$ is the spin $\frac{1}{2}$
irreducible unitary representation of Lorentz group, given by
\begin{equation}
D[R(p)]=\left(
          \begin{array}{cc}
            e^{-i\frac{\phi}{2}} & 0 \\
            0 & e^{i\frac{\phi}{2}} \\
          \end{array}
        \right)\left(
                 \begin{array}{cc}
                   \cos\frac{\theta}{2} & -\sin\frac{\theta}{2} \\
                   \sin\frac{\theta}{2} & \cos\frac{\theta}{2} \\
                 \end{array}
               \right)\label{matrix}
\end{equation}
with
$\hat{\mathbf{p}}=(\sin\theta\cos\phi,\sin\theta\sin\phi,\cos\theta)$.
Note that the spin states and helicity states consider here are
observed within the same inertial reference frame.

For a pure one particle state, we can represent it as expansion of
the spin states, i.e.,
\begin{equation}
|\psi\rangle=\sum_{\sigma=\pm\frac{1}{2}}\int
d^3\mathbf{p}\psi(\sigma,\mathbf{p})|p,\sigma\rangle \label{sstate}
\end{equation}
with the normalized condition
\begin{equation}
 \sum_{\sigma=\pm\frac{1}{2}}\int
d^3\mathbf{p}|\psi(\sigma,\mathbf{p})|^2=1.
\end{equation}
It is noteworthy that this normalized state with a superposition of
various momenta represents a more physical reality since a particle
has no definite momentum in general, although for convenience the
momentum eigenstates are extensively employed in textbooks on high
energy physics and quantum field theory. Then
 the reduced spin
density matrix associated with the above normalized state is
obtained by tracing the momentum degree, i.e.,
\begin{eqnarray}
\rho&=&Tr_\mathbf{p}[|\psi\rangle\langle\psi|]=\int
d^3\mathbf{p}\langle\mathbf{p}|\psi\rangle\langle\psi|\mathbf{p}\rangle
\nonumber\\
&=&\sum_{\sigma,\tilde{\sigma}}\int
d^3\mathbf{p}[\psi(\sigma,\mathbf{p})\psi^*(\tilde{\sigma},\mathbf{p})|\sigma\rangle\langle\tilde{\sigma}|].\label{sdensity}
\end{eqnarray}
Here, we have used the orthonormal relation for the spin states.

Next, as mentioned above, $|\psi\rangle$ can also be expanded by the
helicity states as
\begin{equation}
|\psi\rangle=\sum_{\lambda=\pm\frac{1}{2}}\int
d^3\mathbf{p}\psi'(\lambda,\mathbf{p})|p;\lambda\rangle.
\end{equation}
According to the transformation relation between the spin states and
helicity states Eq.(\ref{TR}), we have
\begin{equation}
\psi'(\lambda,\mathbf{p})=D^{-1}_{\lambda\sigma}[R(p)]\psi(\sigma,\mathbf{p}),
\end{equation}
which follows the reduced helicity density matrix as
\begin{eqnarray}
\rho'&=&\sum_{\lambda\tilde{\lambda}}\int d^3\mathbf{p}
\nonumber\\
&&\{D^{-1}_{\lambda\sigma}[R(p)]\psi(\sigma,\mathbf{p})\psi^*(\tilde{\sigma},\mathbf{p})D_{\tilde{\sigma}\tilde{\lambda}}[R(p)]|\lambda\rangle\langle\tilde{\lambda}|\}.\label{hdensity}\nonumber\\
\end{eqnarray}

As an example, consider a particle prepared with the spin in $z$
direction, i.e., $\psi(-\frac{1}{2},\mathbf{p})=0$, which implies
that the corresponding  spin entropy is zero. However, it does not
mean that the helicity entropy also vanishes for this state. By
Eq.(\ref{matrix}) and Eq.(\ref{hdensity}), the reduced helicity
density matrix can be explicitly written as
\begin{equation}
\rho'=\int d^3\mathbf{p}|\psi(\frac{1}{2},\mathbf{p})|^2\left(
        \begin{array}{cc}
          \frac{1+\cos\theta}{2} & -\frac{\sin\theta}{2} \\
          -\frac{\sin\theta}{2} & \frac{1-\cos\theta}{2} \\
        \end{array}
      \right)
\end{equation}
In particular, for simplicity but without loss of generality, let
$\psi(\frac{1}{2},\mathbf{p})$ be arbitrary except independent of
the angle $\theta$, then the reduced helicity density matrix can be
calculated out as
\begin{equation}
\rho'=\left(
        \begin{array}{cc}
          \frac{1}{2} & -\frac{\pi}{8} \\
          -\frac{\pi}{8} & \frac{1}{2} \\
        \end{array}
      \right),
\end{equation}
whose eigenvalues are easy to obtain as
$\rho'_1=\frac{1}{2}+\frac{\pi}{8}$ and
$\rho'_2=\frac{1}{2}-\frac{\pi}{8}$, respectively. Whence, the
helicity entropy reads
\begin{equation}
S=-[(\frac{1}{2}+\frac{\pi}{8})\log_2(\frac{1}{2}+\frac{\pi}{8})+(\frac{1}{2}-\frac{\pi}{8})\log_2(\frac{1}{2}-\frac{\pi}{8})],\label{entropy}
\end{equation}
which is obviously positive.

Note that, in the case considered above, the helicity entropy does
not depend on the specific form of wave function
$\psi(\frac{1}{2},\mathbf{p})$ in deed. Especially, if we choose
$\psi(\frac{1}{2},\mathbf{p})$ to be a Gaussian, i.e.,
\begin{equation}
\psi(\frac{1}{2},\mathbf{p})=\pi^\frac{3}{4}\tau^{-\frac{3}{2}}e^{-\frac{\mathbf{p}^2}{2\tau^2}},\label{Gaussian}
\end{equation}
which is a minimum uncertainty state, the helicity entropy is
independent of the distribution width parameter $\tau$.

On the other hand, it is noteworthy that the above result does not
show that the helicity entropy always exceeds the corresponding spin
entropy for the same state. In order to confirm this assertion, we
first combine Eq.(\ref{sdensity}) with Eq.(\ref{hdensity}) to obtain
the reduced spin density matrix as
\begin{eqnarray}
\rho&=&\sum_{\sigma\tilde{\sigma}}\int d^3\mathbf{p}
\nonumber\\
&&\{D_{\sigma\lambda}[R(p)]\psi'(\lambda,\mathbf{p})\psi'^*(\tilde{\lambda},\mathbf{p})D^{-1}_{\tilde{\lambda}\tilde{\sigma}}[R(p)]|\sigma\rangle\langle\tilde{\sigma}|\},\label{spiny}\nonumber\\
\end{eqnarray}
if the corresponding reduced helicity density matrix reads
\begin{equation}
\rho'=\sum_{\lambda\tilde{\lambda}}\int
d^3\mathbf{p}[\psi'(\lambda,\mathbf{p})\psi'^*(\tilde{\lambda},\mathbf{p})|\lambda\rangle\langle\tilde{\lambda}|].
\end{equation}
Now consider instead a particle in the eigenstate of helicity with
an eigenvalue $+\frac{1}{2}$ and isotropic momentum distributions
such as Gaussian. Obviously, associated with this state, the
helicity entropy vanishes. However, straightforward but lengthy
calculations lead to the reduced spin density matrix as
\begin{equation}
\rho=\int d^3\mathbf{p}|\psi'(\frac{1}{2},\mathbf{p})|^2\left(
        \begin{array}{cc}
          \frac{1+\cos\theta}{2} & \frac{e^{-i\phi}\sin\theta}{2} \\
          \frac{e^{i\phi}\sin\theta}{2} & \frac{1-\cos\theta}{2} \\
        \end{array}
      \right)=\left(
                \begin{array}{cc}
                  \frac{1}{2} & 0 \\
                  0 & \frac{1}{2} \\
                \end{array}
              \right),
\end{equation}
which follows that the spin entropy takes $1$ in terms of binary
logarithm. Similarly, this result does not rely on the specific
momentum distribution of right handed helicity state except that
isotropy is required here. In addition, this resultant spin entropy
value implies that spin is maximally entangled with momentum for the
prepared isotropic right handed helicity state.

In conclusion, although both the helicity states and spin states can
constitute the basis of Hilbert space for one particle, they have
significantly distinct dependence on momentum degree of freedom and
thus demonstrate remarkably different properties of entanglement.
Thus, even if in high energy physics experiments the prepared state
is a direct product of a function of momentum and a function of
spin, it is not a direct product of momentum and helicity, namely,
the helicity and momentum appear to be entangled for this state,
which means that the measurement of momentum will cause the collapse
of helicity degree of freedom in the inner space of the particle,
and vice versa. This result may be of instructive significance to
both proposal of experiment schemes and analysis of experiment data
related to helicity and spin in high energy physics. In addition, it
is found that both helicity entanglement for a spin eigenstate and
spin entanglement for a right handed or left handed helicity states
do not vanish and their  Von Neumann entropy has no dependence on
the specific form of momentum distribution as long as it is
isotropic.

Stimulating communications on the Shining Star Science Forum in
web(www.changhai.org), especially with Zhangqi Yin are much
appreciated. Work by SH was supported by NSFC(Nos.10235040 and
10421003). SS was supported by LFS of BNU, and HZ was supported in
part by NSFC(No.10533010).


\begin{thebibliography}{99}
\bibitem{Walls}D. F. Walls and G. J. Milburn, Quantum
Optics(Springer-Verlag, Berlin, 1994).
\bibitem{Aspect}A. Aspect, Phys. Rev. Lett. 49: 1804(1982).
\bibitem{Weihs}G. Weihs \emph{et al}, Phys. Rev. Lett. 81: 5039(1998).
\bibitem{Bouwmeester}D. Bouwmeester \emph{et al.}, Nature 390:
575(1997).
\bibitem{Furusawa}A. Furusawa \emph{et al.}, Science 282: 706(1998).
\bibitem{Czachor1}M. Czachor, Phys. Rev. A55: 72(1997).
\bibitem{Peres1}A. Peres and D. R. Terno, J. Mod. Opt. 49:
1255(2002).
\bibitem{Peres2}A. Peres \emph{et al.}, Phys. Rev. Lett. 88: 230402(2002).
\bibitem{Alsing1}P. M. Alsing and G. J. Milburn, quant-ph/0203051.
\bibitem{Gingrich}R. M. Gingrich and C. Adami, Phys. Rev. Lett. 89:
270402(2002).
\bibitem{Peres3}A. Peres and D. R. Terno, J. Mod. Opt. 50:
1165(2003).
\bibitem{Peres4}A. Peres and D. R. Terno, Int. J. Quant. Inf. 1:
225(2003).
\bibitem{Pachos} J. Pachos and E. Solano, Quant. Inf. Comput. 3: 115(2003).
\bibitem{Enk}S. J. van Enk and T. Rudolph, Quant. Inf. Comput. 3: 423(2003).
\bibitem{Czachor2}M. Czachor and M.
Wilczewski, Phys. Rev. A68: 010302(2003).
\bibitem{Bergou}A. J. Bergou \emph{et al.} Phys. Rev. A68:
042102(2003).
\bibitem{Alsing2}P. M. Alsing and G. J. Milburn, Phys. Rev. Lett. 91:
180404(2003).
\bibitem{Peres5}A. Peres and D. R. Terno, Rev. Mod. Phys. 76: 93(2004).
\bibitem{Soo}C. Soo and C. C. Y. Lin, Int. J. Quant. Inf. 2:
183(2004).
\bibitem{Alsing3}P. M. Alsing \emph{et al.}, J. Optics. B6: S834(2004).
\bibitem{Shi1}Y. Shi, Phys. Rev. D70: 105001(2004).
\bibitem{Kim}W. T. Kim and E. J. Son, Phys. Rev. A71: 014102(2005).
\bibitem{Czachor3}M. Czachor, Phys. Rev. Lett. 94: 078901(2005).
\bibitem{FM}I. Fuentes-Schuller and R. B. Mann, Phys. Rev. Lett. 95:
120404(2005).
\bibitem{Caban1}P. Caban and J. Rembielinski, Phys. Rev. A72:
012103(2005).
\bibitem{Kok}P. Kok and S. L. Braunstein, Int. J. Quant. Inf. 4: 119(2006).
\bibitem{Ball}J. L. Ball \emph{et al.}, Phys. Lett. A359: 550(2006).
\bibitem{Lamata1} L. Lamata \emph{et al.}, Phys. Rev. Lett. 97: 250502(2006).
\bibitem{Lamata2}L. Lamata \emph{et al.}, Phys. Rev. A73: 012335(2006).
\bibitem{Jordan1}T. F. Jordan \emph{et al.}, Phys. Rev. A73: 032104(2006).
\bibitem{Alsing4}P. M. Alsing \emph{et al.}, Phys. Rev. A74: 032326(2006).
\bibitem{Shi2}Y. Shi, Phys. Lett. B641: 486(2006).
\bibitem{Caban2}P. Caban and J. Rembielinski, Phys. Rev. A74: 042103
(2006).
\bibitem{Jordan2}T. F. Jordan \emph{et al.}, quant-ph/0608061.
\bibitem{Ling}Y. Ling \emph{et al.}, J. Phys. A40: 9025(2007) .
\bibitem{Adesso}G. Adesso \emph{et al.}, quant-ph/0701074.
\bibitem{He}S. He \emph{et al.}, J. Phys. A40: F857(2007).
\bibitem{Weinberg}S. Weinberg, in Lectures on Particles and Field
Theory, edited by S. Deser and F. W. Ford((Prentice-Hall, Inc.,
Englewood Cliffs, N. J., 1964), vol. II of Lectures delivered at
Brandeis Summer Institute in Theoretical Physics.
\end{thebibliography}
\end{document}